# About solving the Fechner-Stevens problem

*Vasily M. Romanchak*

Belarusian National Technical University, Minsk, Republic of Belarus

**Correspondence:** romancchak@bntu.by

## Abstract

Fechner's and Stevens' laws are the basic psychophysical laws. The problem is that Fechner's and Stevens' laws don't coincide. Researchers have proposed many ways to solve this problem. But attempts to solve the problem continue. In this paper, we prove that the Fechner and Stevens laws are equivalent (coincide up to isomorphism). Therefore, the problem does not exist.

G. Fechner and S. Stevens identified different measurement methods. But only Stevens's definition is the basis of the representative theory of measurement, which developed as a theory of psychophysical measurements. The representative theory has some drawbacks. In particular, there is no built-in mechanism for verifying the adequacy of measurement results. And the question remains, what to do with Fechner's law. Therefore, measurements in psychophysics do not meet strict mathematical criteria. There are even radical proposals to view psychophysics as a pathological science.

The paper considers the solution to the Fechner-Stevens problem (rating method). To this goal, we define two isomorphic algebraic structures. Different algebraic structures correspond to different measurement methods. In the rating method, the two psychophysical laws are equivalent (coincide up to isomorphism) and differ by the measurement methods. Our approach to solving the problem the Fechner-Stevens problem is constructive since it allows one to verify the adequacy of the mathematical model. In this article, we look at an example of subjective measurement using various measurement methods. The example includes a procedure for checking the adequacy.

## Introduction

Fechner (1860) was the first to use subjective measurements method. Fechner's method made it possible to formulate the psychophysical law of the form $u = \lambda_1 \ln(q)$, where $u$ is an objective value, $q$ is a subjective value, $\lambda_1$ is a constant, $\lambda_1 > 0$. Fechner's method based on counting the number of "just noticeable differences" between sensations associated with pairs of stimuli, such as two sounds of different intensities. Thus, it is possible to determine the number of noticeable differences between any two pairs of stimuli (the difference in values).

We will consider the objects $\omega_1, \omega_2, ..., \omega_n$. Fechner's law for the difference of subjective values has the form

$$u_i - u_j = \lambda_1 \ln(q_i / q_j), \tag{1}$$

where $i, j = 1, 2, ..., n$, $n > 2$, $u_i = u(\omega_i)$, $q_i = q(\omega_i)$, $q_i$ are the objective values, $q_i > 0$; $u_i$ are subjective values.

Stevens (1957) disagreed with Fechner and defined measurement as "the assignment of numerals to objects and events according to a rule." Stevens proposed to replace Fechner's law with a psychophysical law of the form $v = cq^{\lambda_2}$, where $c, \lambda_2$ are constants, $c > 0, \lambda_2 > 0$. Stevens law for the ratio of subjective values has the form

$$\ln(v_i / v_j) = \lambda_2 \ln(q_i / q_j), \tag{2}$$

where $i, j = 1, 2, ..., n$, $\lambda_2$ is a constant, $v_i = v(\omega_i)$, $q_i = q(\omega_j)$, $q_i$ are the objective values, $q_i > 0$; $v_i$ are the subjective values, $v_i > 0$. We assume that the values of $u_i$ are determined up to an arbitrary additive constant, and the values $v_i$ are determined up to an arbitrary multiplicative constant.

Two different psychophysical laws lead to different subjective values. Various modifications of the laws of Fechner and Stevens (Cook, 1967), (Ekman,1964), (Krueger, 1989), (Norwich, Wong,1997) explain this problem in different ways. The history of the two laws and the current state of the issue can be found in (Grondin, 2016), (Lubashevsky, 2019).

Lubashevsky (2019) came to the conclusion that there is still a complex problem in psychophysics, which consists in the fact that the basic psychophysical laws contradict each other. The contradiction is that each of the laws describes a person's reaction to the sensation of external stimuli, but their functional forms are different. The reconciliation of these two laws is the subject of a long debate, but the participants in the debate did not come to a common opinion. Nevertheless, Stevens ' definition of measurement is the basis of representative measurement theory.

**Analysis of problem**

Barzilai (2010) presented a new measurement theory and offered to correct fundamental errors in measurement theory. He emphasized that that a mathematical operation is a valid element of a model only if it is a homomorphic image of an empirical operation. Other operations do not apply to scales and scale values. For example, let the measurement result be the ratio of values: $A_2 / A_1 = 2$. Suppose that $A_1 = 1$, then $A_2 = 2$. Thus, we have defined an operation that is adequate to the measurement results.

An error occurs if we assume that $A_1$ and $A_2$ values on the scale of relations and we find, for example, the difference: $A_2 - A_1$. But the difference is not defined, since there is no empirical justification for the operation. And the expression $C(A_2 - A_1)$, where $C$ is a constant, is also undefined. Barzilai (2007) showed that the existing methods of subjective measurement based on similar errors, which have led to many methodologies that produce meaningless numbers. He suggested correcting fundamental errors in measurement theory, decision theory, and mathematical psychology.

The rating method is proposed to correct the Fechner-Stevens problem. The rating is an invariant of two isomorphic algebraic structures with different measurement operations. The main methods of subjective measurement are equivalent, it follows from the definition of the rating. Psychophysical laws differ only in the measurement operation. This solution of the Fechner-Stevens problem is constructive as it contains a possibility of experimental verification of measurement results.

## Fechner's and Stevens' Laws

We find the difference of values $(u_i - u_j)$ using Fechner's law (1), the ratio of values $(v_i / v_j)$ according to Stevens ' law (2). In this case, the measurement results (difference and ratio of values) are related to each other by the formula

$$(u_i - u_j) = \lambda \ln(v_i / v_j), \qquad (3)$$

where $i, j = 1, 2, \ldots, n$; $u_i$ and $v_i$ are subjective values in the Fechner and Stevens law, $\lambda = \lambda_1 / \lambda_2$; $\lambda_1, \lambda_2$ are constants. Equality (3) follows directly from Fechner's (1) and Stevens' (2) laws.

The function $u = \lambda \ln(v)$ is an isomorphism which translates division of positive real numbers into a subtraction of real numbers. Algebraists do not distinguish between isomorphic structures Equality (3) means that the isomorphism translates measurement results $(v_i / v_j)$ into $(u_i - u_j)$ . [1]. Such measurement results are called equivalent. For illustration, consider a special case when the size of the measured objects changes uniformly.

For example, in Table 1 there are circles, where the area of each circle after the first one is found by multiplying the area of the previous one by a fixed number 2, i. e. $v_{i+1} / v_j = 2$, i = 1, 2, \ldots, 5. Let's assign an ordinal number to each object, $u_i = i$ and find the measurement result by the formula

$$u_i - u_j = i - j.$$

The second method of measurement is performed using the relations specified in the condition. Then for the measurement result we obtain the expression

$$v_i / v_j = 2^i / 2^j,$$

where $i, j = 1, 2, \ldots, 6$. The measurement results are equivalent, since the isomorphism (3) is defined, where $\lambda = 1 / \ln(2)$.

Table 1

Circles (area varies uniformly).

| $i$ | 1 | 2 | 3 | 4 | 5 | 6 |
|---|---|---|---|---|---|---|
| $\omega_i$ | 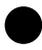 | 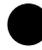 | 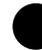 | 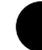 | 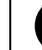 | 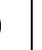 |

It follows from the example that measurements can be made both by subjective and objective methods, and Fechner's method can be applied to any objects whose size varies uniformly. For the two methods of measurement, we can obtain values (Table 2.), for which the isomorphism $u = \lambda\ln(v)$, where $\lambda = 1 / \ln(2)$, establishes a one-to-one correspondence between the values.

Table 2

Values, obtained by two methods.

| $u_i$ | 1 | 2 | 3 | 4 | 5 | 6 |
|---|---|---|---|---|---|---|
| $v_i$ | 2 | $2^2$ | $2^3$ | $2^4$ | $2^5$ | $2^6$ |

The measurement results obtained by the Fechner and Stevens method are equivalent. If the ratio of values in Stevens' law is known, then the difference of values in Fechner's law according to formula (3) is determined, and vice versa. The equivalence of the measurement results means the solution of the Fechner-Stevens problem.

## Rating definition

Let the measurement result be equal to either the difference between the values $(u_i - u_j)$ or the ratio of the values $v_i / v_j$. Thereby, two main measurement methods are defined. Two measurement methods can be considered simultaneously. Let us replace equality (3) by two expressions

$$R_{ij} = \lambda_2(u_i - u_j), \tag{4}$$

$$R_{ij} = \lambda_1 \ln(v_i / v_j), \tag{5}$$

where $i, j = 1, 2, …, n$. The mappings (4) and (5) are called the rating. The rating does not depend on the measurement method (4) or (5). Fechner's and Stevens' laws experimentally confirm the existence of two measurement methods (4) and (5)

The adequacy of the measurement results can be established using the rating definition. The respondent may be mistaken or misleading. To check the respondent's answers, we can use two measurement methods independently of each other. The next section details an example of real data analysis.

**Example.**

Let the area of six circles $\omega_1, \ \omega_{2, …,} \ \omega_6$ (Table 1) are measured subjectively. The respondent's answers to the questions are presented in Table 3.

Table 3.

Measurement results.

| $u_i - u_4$ | -7 | -4 | -2 | 0 | 2 | 4 |
|---|---|---|---|---|---|---|
| $v_i / v_1$ | 1.00 | 1.20 | 1.50 | 2.00 | 2.50 | 3.00 |

The measurement results in the first row are obtained using the semantic differential method. The respondent compares all objects to a fixed one and specifies an integer between -8 and 8. The number should correspond to the degree of superiority of one object over another. For example, the number -7 means that the first object is seven units less than the fixed (fourth) object.

Then the measurement results are obtained by the second method. The respondents' answers are in the second row of Table 3. The respondent sequentially selects all objects and compares with a fixed object. We assume, for example, that the fixed object is first. The respondent indicates how many times the selected object is larger than the first object. For example, the respondent believes that the third object is one and a half times larger than the first. The respondent's answer is in the second row of the fourth column of the table (3). Let be

$$R_{i4}(1) = u_i - u_4$$

$$R_{i1}(2) = \ln(v_i / v_1)$$

where $R_{i4}(1)$ is the rating, obtained by the first measurement method (4), $R_{i1}$ (2) is rating obtained by the second measurement method (5). Then $R_{i1}$ (1) = $R_{i4} - R_{14}$. Rating values $R_{i1}(1)$ and $R_{i1}(2)$ are in the first and second row of Table 4.

Table 4

Rating values.

| $R(1)$ | 0.0 | 3.0 | 5.0 | 7.0 | 9.0 | 11.0 |
|--------|------|------|------|------|------|------|
| $R(2)$ | 0.00 | 0.18 | 0.41 | 0.69 | 0.92 | 1.10 |

The regression equation for the data in Table 4 is $r_1 = b_0 + b_1 r_2$, where $b_1 = 0.106$, $b_0 = -0.067$; $b_1$ coefficient is statistically significant ($t$-test, achievable significance level $p = 0.00008$); the coefficient $b_0$ not statistically significant ($t$-test, achievable significance level $p = 0.20$). Therefore, we can accept the hypothesis of mutual adequacy of the rating values $R(1)$ and $R(2)$. The graph of the regression equation, Fig. 1, confirms this conclusion.

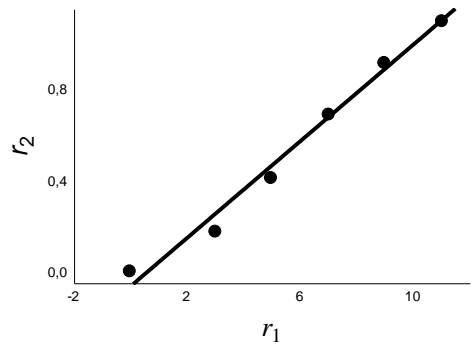

*Figurer* 1. Linear regression

Such a preliminary analysis, despite the small amount of statistical data, allows for individual control of each respondent and avoids gross errors when testing a group of respondents.

## Conclusion

The article shows that psychophysical measurements can be carried out by two methods. Moreover, the results of measurements performed by the two methods are equivalent (coincide up to isomorphism). The mathematical model of measurement is two isomorphic algebraic structures. The structure consists of a set of real numbers with measurement operation, subtraction or division. Fechner's and Stevens' laws confirm the adequacy of the mathematical model of measurement. The article presents an example of an application, which could be used in psychological measurements. The example includes a procedure for checking the equivalence of measurement results.